\documentclass{mem}
\usepackage{natbib}\usepackage{txfonts}\usepackage{balance}
\usepackage{graphicx}
\usepackage[breaklinks]{hyperref}
\idline{75}{282}
\begin{document}
\def\teff{$T\rm_{eff }$}
\def\kms{$\mathrm {km s}^{-1}$}

\title{
MAGIC extragalactic highlights from a MeV perspective
}

   \subtitle{}

\author{
E. Prandini\inst{1} for the MAGIC Collaboration
          }

\institute{
Dipartimento di Fisica e Astronomia G. Galilei, Universit\'a degli Studi di Padova, Italy \&
I.N.F.N. Sez. di Padova
\email{elisa.prandini@unipd.it}
}

\authorrunning{Prandini}

\titlerunning{MAGIC extragalactic highlights}

\abstract{
In the past fifteen years, the way to study TeV gamma-ray emitters changed drastically. The detection-based approach aimed at populating the TeV gamma-ray sky evolved into a physics-driven one, with the ambitious objective of understanding the mechanisms responsible for the emission and their environments. The synergic collaboration between instruments operating in different electromagnetic bands and with different messengers is therefore crucial. In this talk, I will report highlights on extragalactic physics studies achieved with the MAGIC telescopes, with special emphasis on the MeV-TeV connection.
}
\maketitle{}

%
%

\section{Introduction}
The study of the most energetic photons from galactic and extragalactic emitters, the  very high-energy photons (VHE, E$>$100\,GeV) is one of the most recent disciplines in observational astronomy. 

The current generation of Imaging Atmospheric  Cherenkov Telescopes (IACTs) includes two arrays of telescopes located in the Northern hemisphere, namely  the Major Atmospheric Gamma-ray Imaging Cherenkov (MAGIC) and the Very Energetic Radiation Imaging Telescope Array System (VERITAS), and one array in the Southern hemisphere, the High Energy Stereoscopic System (H.E.S.S.). 

\subsection{The MAGIC telescopes}
Since 2004, the MAGIC telescopes observe the Northern sky from the Canary island of La Palma, at $\sim$2200\,m above the sea level.
As all IACTs, MAGIC collects the faint, rapid (few nanoseconds) Cherenkov flashes of optical-UV light emitted from the shower of particles induced by a VHE gamma-ray entering the Earth atmosphere.

Main source of background is constituted by proton-induced atmospheric showers. All optical-UV sources of night photons, such as the moon, disturb the observations decreasing the overall sensitivity \citep{MAGIC_moon2017}.

The performance of the MAGIC telescopes was evaluated with deep observations of the Crab Nebula, as standard candle \citep{MAGIC_perf2016}.
The two key features of the MAGIC telescopes with respect to other IACTs are a low energy threshold of $\sim$60\,GeV and a fast positioning  system, essential in case of fast alerts such as gamma-ray bursts (GRBs) and neutrino alerts.
The MAGIC science studies are divided into the following working groups: Galactic, Extragalactic, Transient, and Fundamental Physics.

Due to its location, low energy threshold, and fast repositioning MAGIC is best suited for studies involving extragalactic sources and transient events.

\subsection{The extra-galactic MAGIC sky}
The extragalactic sky at VHE gamma rays is completely dominated by jetted Active Galactic Nuclei (AGNs): only three sources out of the $\sim$80 sources listed in the TeV catalog (\url{http://tevcat2.uchicago.edu}) are non-AGNs. These sources are the two starburst galaxies M~82 and NGC~253, and the gamma ray burst GRB~190114C.

\begin{figure}
    \centering
    \includegraphics[width=0.49\textwidth]{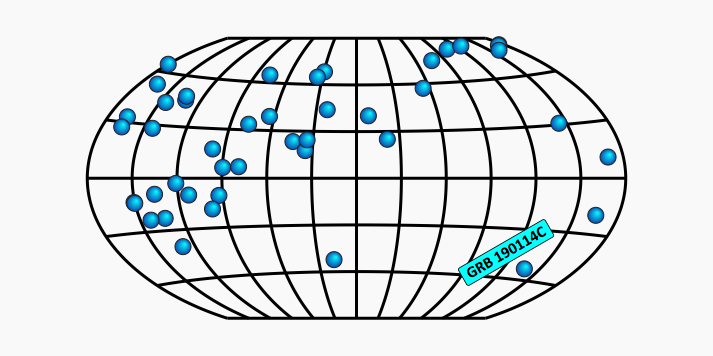}
    \caption{Map of the sources detected as VHE gamma-ray emitters with the MAGIC telescopes. Adapted from \url{http://tevcat2.uchicago.edu}.}
    \label{fig:MAGIC_map}
\end{figure}

In its first 16 years of operations, the MAGIC telescopes detected  significant signals from more than  half of the known extragalactic TeV emitters. The 44 extragalactic objects detected with MAGIC as of April 2019 are illustrated in Fig.~\ref{fig:MAGIC_map}. In the Figure the GRB~190114C (ATel~\#12390), is highlighted since it is a very important result for the MAGIC collaboration and, more in general, for the growing multi-messenger astrophysical community. 
However, at the moment the analysis of the MAGIC data of this exceptional event is ongoing and no information is still available.
All the other MAGIC detected sources belong to the AGN class. 

In the first years of MAGIC operation the research field was still in its pioneering stage, and the main aim of the observations was that of populating the TeV sky.  
The discipline is now matured into more physics-driven studies, targeted at understanding the mechanisms responsible for the emission and their environments. To these purposes, the role of VHE gamma ray observations is the evaluation of the time variability of the emission and the measure of precise spectra and their cut-offs.
The study of the interplay between VHE gamma rays and photons at shorter wavelengths is fundamental to achieve the physics goal and coordinated, broadband observation campaigns are the becoming the standard in the field. Moreover, the  connection with other messengers, like neutrinos, plays a crucial role and should be taken into consideration.

For this contribution, three highlights were selected with a special attention on the possible MeV--TeV connection.

\section{Highlight Results}
\subsection{Flat Spectrum Radio Quasars}
Blazars, jetted-AGNs with a jet pointing towards the observer, are divided into two main classes depending on the presence or not of broad lines in their optical spectrum. BL Lac objects feature no or very weak emission lines while flat spectrum radio quasars (FSRQs) are characterized by strong optical lines due to a dense cloud region distributed between the jet and the disk. On average, FSRQs are more luminous and they are detected up to redshift above 2, while known BL Lac objects are fainter and their distance is limited (mostly below 1 for gamma-ray detected objects, see for example the 3LAC catalog \citep{3lac}. In both cases, the spectral energy distribution (SED) is dominated by a double peaked, non thermal-emission from the jet.

The large majority of AGNs detected by IACTs in the VHE gamma-ray range belongs to the BL Lac class that despite the limited luminosity is characterized by an SED shifted towards higher energies with respect to the FSRQ case. 
In FSRQ, the second SED peak usually lies in the MeV to GeV range, making them ideal targets for MeV instruments. In particular, the MeV range is essential for the study of distant FSRQs, see \citet{sbarrato2015}.

Only 8 FSRQs populate the VHE extragalactic sky. They are listed in Table~\ref{tab:fsrq}. Due to the second SED peak frequency, well below the VHE range, FSRQs are usually detected by IACTs only at low energies ($\sim$100\,GeV) and during bright flares. The only source detected at VHE also in its persistent state is PKS 1510-089 \citep{magic2018a}.

\begin{table}[]
    \centering
    \begin{tabular}{lccc}
    \hline
    \hline
    Source & z & Discoverer & Year \\
    \hline
    \hline
    B 0218+367  & 0.944 & MAGIC & 2014 \\
    PKS 1441+25 & 0.939 & MAGIC & 2015 \\ 
    TON 599     & 0.720 & MAGIC & 2017 \\
    3C 279      & 0.536 & MAGIC & 2006 \\
    S4 0954+65* & 0.356? & MAGIC & 2015 \\
    PKS 1222+216 & 0.432 & MAGIC & 2010 \\
    PKS 1510-089 & 0.361 & HESS & 2009 \\
    PKS 0736+017 & 0.189 & HESS & 2016 \\
    \hline
    \hline
    \end{tabular}
    \caption{List of FSRQs detected by current generation of IACTs. 
       The classification of S4 0954+65, marked with $*$, is uncertain.}
    \label{tab:fsrq}
\end{table}

Thanks to the good performance at low energies \citep{MAGIC_perf2016}, FSRQs are ideal targets for MAGIC. Since these objects show fast, day-scale variability at all bands, a quick reaction to alerts is essential and was the key aspect for many MAGIC discoveries. 

The main open question adressed with VHE observations is related to the location of the emitting region in FSRQs, which is strongly connected to the VHE spectral cut-off and to the variability timescale. 
According to the standard scenario of non-thermal emission from FSRQs, a short, minute-scale, variability implies a strong self-absorption already  at few tens of GeV, while a day-scale variability is compatible with a non-self absortion scenario.
In MAGIC observations of the source PKS~1222+216,  the minutes-scale VHE variability seems in contradiction with the observed spectral shape \citep{magic2011}. On the other hands, in other objects such as the recently discovered TON~599 (MAGIC Coll. in prep.) the standard scenario depicted above is compatible with MAGIC observations. Further, multi-wavelength observations of FSRQs both during the persistent and the flaring states are needed to test the standard scenario of emission in this class of objects.

\subsection{The {\it Neutrino blazar}} 
TXS~0506+056 is a gamma-ray emitting blazar located at redshift 0.336 \citep{paiano2018}.
This object was only one of the many gamma-ray emitting blazars detected by {\it Fermi}-LAT at gamma rays, at least until the 22nd September 2017, when IceCube reported the detection of a neutrino from a region compatible with that of TXS~0506+056 (GCN/AMON Notice dated 22 September 2017 20:55:13 UTC).
The alert was immediately followed by multi-wavelength and multi-messenger observations summarized in \citet{icecube2018}. The analysis of {\it Fermi}-LAT data collected after the alert and in the precedent years revealed that TXS~0506+650 was in an active gamma-ray state since several months at the moment of the alert. Moreover, the emission reaches the VHE gamma-ray band, as announced by MAGIC and confirmed by the VERITAS collaboration.

This exceptional event officially marked the beginning of the era of extragalactic multi-messenger astronomy with neutrinos. Every messenger, as well as every band of the electromagnetic spectrum, carries important information needed to unveil the main mechanisms at work in jetted AGNs.
As detailed in  \citet{ojha2019}, the MeV band (including polarization) is an excellent proxy for photo-hadronic processes in blazar jets, that are expected to produce neutrino counterparts.

Sub-TeV observations are also crucial to probe the emitting mechanisms, as discussed in \citet{magic2018b}. In that paper, the broad-band SED including MAGIC sub-TeV spectra both during the quiescent state of the source and during its flaring state are interpreted in the framework of a novel one-zone lepto-hadronic model. According to this  model, electrons and protons co-accelerated in the jet interact with external photons coming from a slow-moving plasma layer surrounding the faster jet spine.

\begin{figure}
    \centering
    \includegraphics[width=0.49\textwidth]{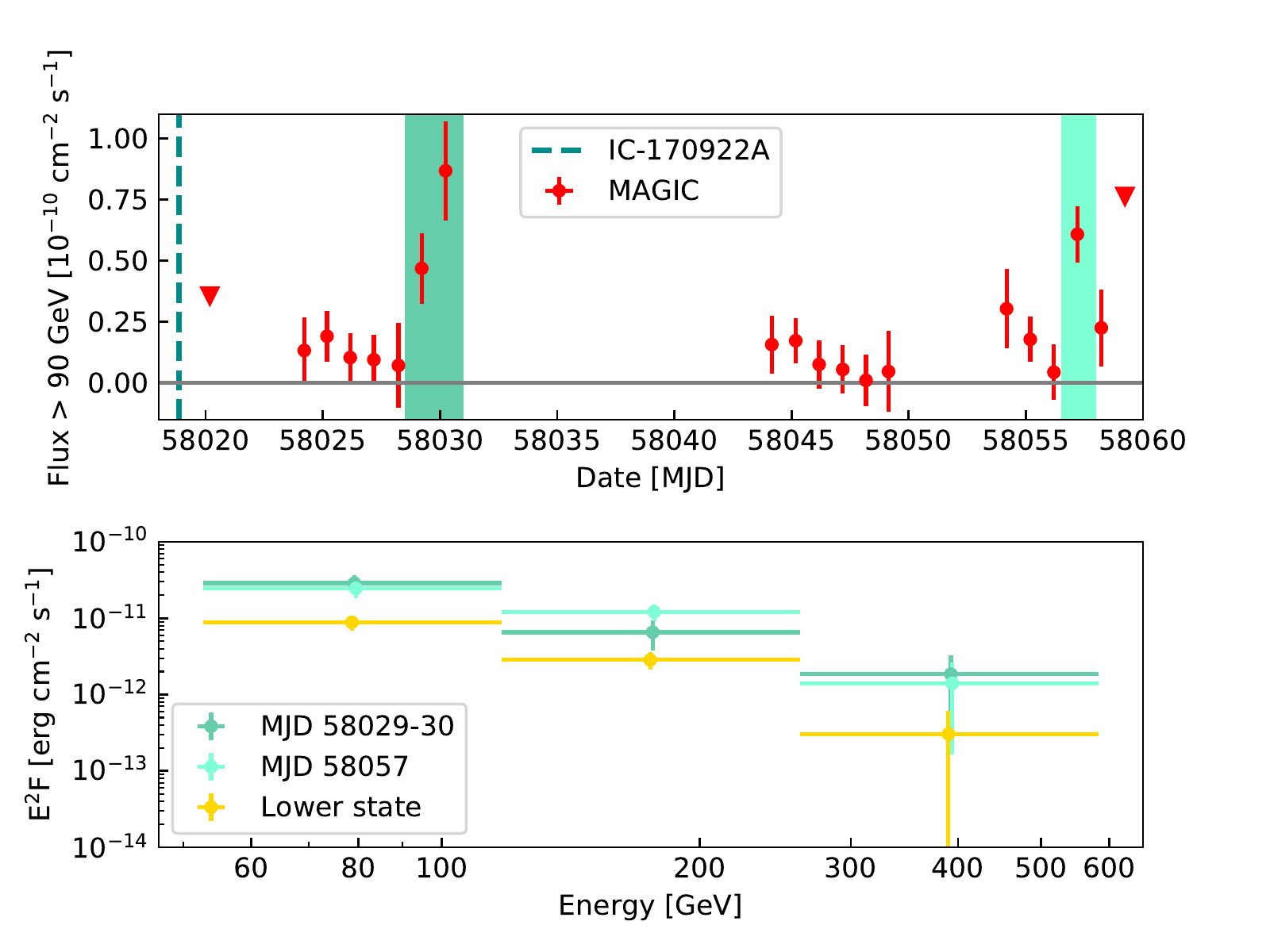}
    \caption{Upper panel: MAGIC light curve above 300\,GeV starting from the neutrino alert (vertical dashed line). Two flaring states were detected (green areas). Lower panel: differential energy spectra at VHE as measured by MAGIC during the two flares reported above (MJD~58029-30 and MJD~58057) and during the low/quescent state. Figure from \citet{magic2018b}.}
    \label{fig:TXS_MAGIC}
\end{figure}

Figure~\ref{fig:TXS_MAGIC} illustrates the results of MAGIC observations. In the light curve (upper panel) two flaring events are highlighted, at MJD~58029-30 and MJD~58057 respectively. The corresponding differential energy spectra are represented in the lower panel (green points), together with the averaged emission obtained from the low state/quiescent nights. Interestingly, the slope of the MAGIC spectrum does not show strong evidence for variability during the different flux states.

Main outcome of the lepto-hadronic model proposed is that a non-negligible contribution in the SED arises from cascade emission induced by protons, most notably in the hard X-ray and VHE gamma-ray bands. 
The maximum energy of protons inferred  from the model is consistent with an important contribution to the flux of ultra high-energy cosmic rays from protons (and heavier nuclei) accelerated in the jet region.

\subsection{Extreme blazars} 
Extreme blazars are a sub-class of blazars  featuring the first SED peak, i.e. the synchrotron peak, above 10$^{17}$\,Hz \citep{costamante2001}.
The second SED peak is therefore possibly shifted in the sub-TeV range. This at least for a sub-sample of objects, as discussed in \citet{foffano2019}.
TeV blazars are interesting targets for MeV observations as their synchrotron peak could lie in this band \citep{deangelis2018}. 

The TeV-peaking SED of the extreme blazar 1ES~0229+200 attracted a number of interesting studies on various subjects. Namely: the study of particle acceleration mechanisms in such extreme objects, e.g. \citet{kaufmann2011}; estimate of new limits on the extragalactic background light \citep{aharonian2007,magic19}, intergalactic magnetic field \citep{vovk2012}, and for axion-like particles \citep{galanti2018};  test for the hadron beam scenario \citep{tavecchio2018}; probe for cosmology and ultra-high-energy cosmic rays \citep{tavecchio2015}; study of Lorentz invariance violation \citep{tavecchio2016}.
Therefore, the impact of the observation and characterization of these extreme blazars for the astrophysical  and cosmological community is very broad. The field, however, suffers from the still quite limited number of objects belonging to this category. 

\begin{figure}
    \centering
    \includegraphics[width=0.49\textwidth]{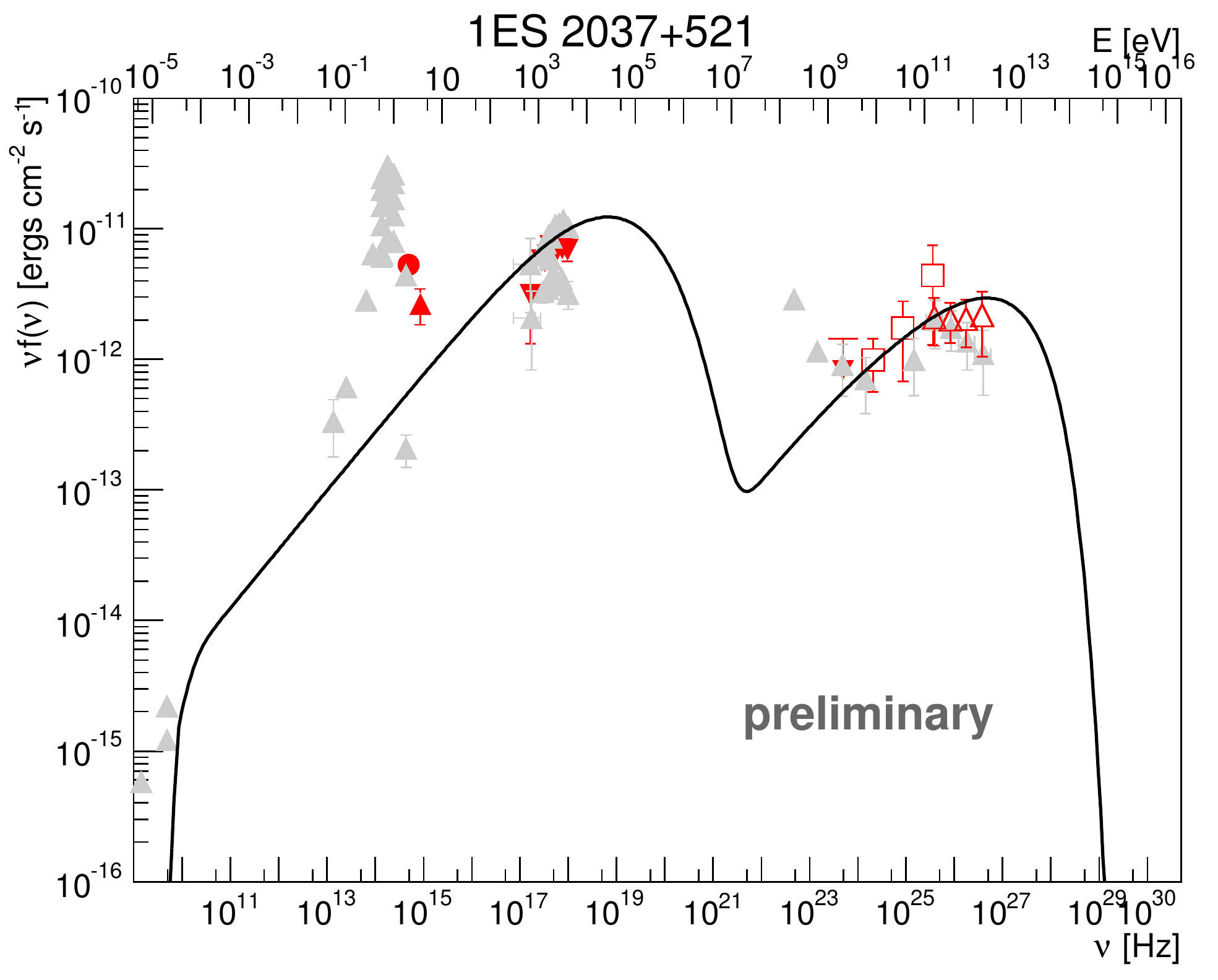}
    \caption{Multi-band SED of 1ES~2037+521 during MAGIC observations in 2016 (red), including {\it Swift}-UVOT, {\it Swift}-XRT, {\it Fermi}-LAT, and MAGIC data, along with  
    archival data (gray). The black curve represents the SSC model \citep{asano2014} fitting the data.}
    \label{fig:SED_2037}
\end{figure}

With the aim of increasing the number of TeV-detected extreme blazars, the MAGIC collaboration started a multi-year observational campaign on several objects. In addition, deep observation of 1ES~0229+200, the prototype of hard-TeV extreme blazars, was performed. Table~\ref{tab:list_ehbl} lists the main results of this campaign: MAGIC observed 11 sources from 2010 to 2018, and for five sources detected a clear signal at VHE gamma rays. For one source, RGB~J2042+244, a strong hint of signal was found.
The SEDs were modeled with a one-zone synchrotron self-Compton (SSC) model, following \citet{asano2014}. The main result is that an extremely low magnetization is required to successfully fit the data. An example of model is displayed in Fig.~\ref{fig:SED_2037}. The model parameters for this and all the other sources listed in Table~\ref{tab:list_ehbl} are described in two MAGIC collaboration papers in preparation.

\begin{table}[]
    \centering
    \begin{tabular}{lcc}
         \hline
     \hline
     & & MAGIC \\
      Source & z & detection\\
     \hline
     \hline
    TXS 0210+515 & 0.049 & Y \\ 
    1ES 2037+521 & 0.053 & Y \\
    PGC 2402248 & 0.065 & Y \\
    BZB J0809+3455 & 0.083 & N \\
    RGB J2042+244 & 0.104 & Hint \\
    1ES 1426+428 & 0.129 & Y \\
    RGB J2313+147 & 0.163  & N \\ 
    1ES 0927+500 & 0.187 & N \\
    RBS 0723 & 0.198 & Y \\
    RBS 0921 & 0.236 & N \\
    TXS~0637-128 & unknown & N \\
    \hline
     \hline
    \end{tabular}
    \caption{Extreme blazars observed with the MAGIC telescopes from 2010 to 2018 ordered according to their distance.}
    \label{tab:list_ehbl}
\end{table}

\section{Conclusions}
In this exciting moment for the high-energy astrophysical community, with the recent birth of multi-messenger extragalactic astronomy with neutrinos, MAGIC has a leading role in the study of VHE gamma-ray emissions from extragalactic sources. 
The photon-neutrino connection from blazars, as well as the characterization of the broadband emission from FSRQs and from extreme blazars presented here are some examples of common interest for the TeV and the MeV communities.

%
%
{\tiny \emph{Acknowledgements:} We would like to thank the Instituto de Astrof\'{\i}sica de Canarias for the excellent working conditions at the Observatorio del Roque de los Muchachos in La Palma. The financial support of the German BMBF and MPG, the Italian INFN and INAF, the Swiss National Fund SNF, the ERDF under the Spanish MINECO (FPA2015-69818-P, FPA2012-36668, FPA2015-68378-P, FPA2015-69210-C6-2-R, FPA2015-69210-C6-4-R, FPA2015-69210-C6-6-R, AYA2015-71042-P, AYA2016-76012-C3-1-P, ESP2015-71662-C2-2-P, FPA2017‐90566‐REDC), the Indian Department of Atomic Energy, the Japanese JSPS and MEXT and the Bulgarian Ministry of Education and Science, National RI Roadmap Project DO1-153/28.08.2018 is gratefully acknowledged. This work was also supported by the Spanish Centro de Excelencia ``Severo Ochoa'' SEV-2016-0588 and SEV-2015-0548, and Unidad de Excelencia ``Mar\'{\i}a de Maeztu'' MDM-2014-0369, by the Croatian Science Foundation (HrZZ) Project IP-2016-06-9782 and the University of Rijeka Project 13.12.1.3.02, by the DFG Collaborative Research Centers SFB823/C4 and SFB876/C3, the Polish National Research Centre grant UMO-2016/22/M/ST9/00382 and by the Brazilian MCTIC, CNPq and FAPERJ.}

\bibliographystyle{aa}

\end{document}